# Pseudogap in a crystalline insulator doped by disordered metals

Sae Hee Ryu[1,4], Minjae Huh[1,2,4], Do Yun Park[1,4], Chris Jozwiak[3], Eli Rotenberg[3], Aaron Bostwick[3] & Keun Su Kim[1*]

[1]Department of Physics, College of Science, Yonsei University, Seoul, Korea.
[2]Department of Physics, Pohang University of Science and Technology, Pohang, Korea.
[3]Advanced Light Source, E. O. Lawrence Berkeley National Laboratory, Berkeley, CA, USA.
[4]These authors contributed equally: Sae Hee Ryu, Minjae Huh, Do Yun Park.
*e-mail: keunsukim@yonsei.ac.kr

**A key to understand how electrons behave in crystalline solids is the band structure that connects the energy of electron waves to their wavenumber ($k$). Even in the phase of matter with only short-range order (liquid or amorphous solid), the coherent part of electron waves still possesses a band structure. Theoretical models for the band structure of liquid metals were formulated more than 5 decades ago[1-15], but thus far, bandstructure renormalization and pseudogap induced by resonance scattering have remained unobserved. Here, we report the observation of this unusual band structure at the interface of a crystalline insulator (black phosphorus) and disordered dopants (alkali metals). We find that a conventional parabolic band structure of free electrons bends back towards zero $k$ with the pseudogap of 30–240 meV from the Fermi level. This is $k$ renormalization caused by resonance scattering that leads to the formation of quasi-bound states in the scattering potential of alkali-metal ions. The depth of this potential tuned by different kinds of alkali metal (Na, K, Rb, and Cs) allows to classify the pseudogap of $p$-wave and $d$-wave resonance. Our results may provide a clue to the puzzling spectrum of various crystalline insulators doped by disordered dopants[16-20], such as the waterfall dispersion in cuprates.**

In 1960s, there was a series of pioneering theoretical works for the band structure of liquid metals[1-9]. Like glassy materials have a well-defined complex refractive index, electron waves under the influence of multiple scattering acquire a complex wavenumber shift $\Delta k$ whose magnitude is sizable at resonance. As shown in Fig. 1a, the band structure of free electrons ($E \sim k^2$) is distorted by the real part of $\Delta k$ [Re($\Delta k$)] in an unusual sinusoidal form that has no counterpart in crystalline solids. The corresponding imaginary part of $\Delta k$ [Im($\Delta k$)] represents a spread of $k$ related to the formation of quasi-bound states (QBS) around liquid ions. In the density of states (inset in Fig. 1a), a local minimum is formed at a resonance energy ($E_r$)[5,9-12], which is the "pseudogap" coined by Mott[13,14]. This backward-bending band dispersion with pseudogap should be observable by means of angle-resolved photoemission spectroscopy (ARPES). ARPES has been used to study the electronic structure of a melting lead monolayer on Cu(111)[21] and Si(111)[22]. However, the characteristic $k$ renormalizations and pseudogap induced by resonance scattering depicted in Fig. 1a, despite their fundamental importance, have remained unobserved experimentally.

A key idea in our experiments is to study the interface of disordered dopants (alkali metals) and a crystalline insulator (black phosphorus)[23-26]. Black phosphorus is a layered material, in which the honeycomb lattice of P atoms is modulated to form an array of zigzag ridges and valleys (Fig. 1b). The distribution of alkali metals on black phosphorus was reported to show the radial (and anisotropic) structure factor[25], which is reproduced by structural simulations in Fig. 1c (see Extended Data Fig. 1 for key assumptions). This radial shape is the key feature of liquid or glassy phases[27,28], and reflects the presence of the mean interatomic distance (or short-range order) generic to dopants randomly distributed under the repulsive interaction. Each alkali metal donates its valence electron to black phosphorus, and the doped electrons are mostly populated in the topmost layer[23,24]. Then, doped electrons are subject to multiple scattering by the potential of dopant ions with only short-range order, which is the situation required to have resonance effects in theoretical models (Fig. 1a)[1-12]. Furthermore, the low diffusion barrier of alkali metals on black phosphorus[26] allows us to systematically trace the evolution of ARPES data by varying the density of dopants at low temperature.

**Backbending dispersion and pseudogap**

In the simple picture of surface doping that has been widely accepted so far, the conduction bands in the topmost layer of black phosphorus adjacent to alkali-metal ions shift below the Fermi energy $E_F$. Figure 1d shows constant-energy contours at $E_F$ (the Fermi surface)[23,24] expected for surface-doped black phosphorus at the electron density $n_e$ of $1.0 \times 10^{14}$ cm$^{-2}$. There is a large oval-shape contour centred at the $\Gamma$ point (labelled C1) with a pair of small pockets separated along the $y$ axis (labelled C2). The energy dispersion of C1 and C2 bands is shown in Fig. 2a-c along 3 high-symmetry directions ($k_y$, $k_s$, and $k_x$ as indicated in Fig. 1d), respectively. Owing to the crystal symmetry of black phosphorus, the band dispersion of C1 changes from quadratic in $k_y$ (Fig. 2a) to linear in $k_x$ (Fig. 2c)[23]. A set of $k$ at $E_F$ (or $k_F$) in every in-plane direction constitutes the oval-shape Fermi contour of C1 in Fig. 1d.

Figure 2d-o displays a series of ARPES data taken from bulk black phosphorus whose surface is doped by different kinds of alkali metal (Na, K, Rb, and Cs), as marked at the bottom right of each panel. In data for Na taken along $k_y$ (Fig. 2d), there is a clear signature of C2 bands at ±0.6 Å$^{-1}$, as expected in Fig. 2a. In contrast, a striking deviation from that expected in Fig. 2a is observed for the C1 band. Unlike the sharp Fermi-Dirac cut-off of C2 bands that can be used as an internal reference of $E_F$, the C1 band shows a gaplike feature with the magnitude of at least 0.2 eV from $E_F$. More noteworthy is that the band dispersion of C1 bends back at around −0.2 eV towards $k = 0$ or the $\Gamma$ point with a weak intensity. This backward-bending dispersion of the C1 band can be seen more clearly in $k_s$ (Fig. 2e) with a diminishing intensity towards $E_F$, and intriguingly even reaches close to $k = 0$ in $k_x$ (Fig. 2f).

On the other hand, a difference from the Na case is observed for black phosphorus doped by K, Rb, and Cs. In data taken along $k_y$ (Fig. 2g,j,m), there is commonly a pair of C2 bands crossing $E_F$, as observed in Fig. 2d. However, the C1 band shows a well-defined energy gap with no sign of the backward-bending band dispersion, as the ARPES intensity diminishes more abruptly towards $E_F$. The magnitude of energy gaps is about 65–70 meV for K and Rb, and about 30 meV for Cs, which are clearly smaller than 200 meV in the Na case (Fig. 2d). A series of ARPES data taken along $k_s$ (Fig. 2h,k,n) and along $k_x$ (Fig. 2i,l,o) consistently shows the well-defined energy gap with nearly the same magnitudes within the range of ±10 meV, which indicates the isotropy of energy gap (Extended Data Fig. 3 for constant-energy maps). This signature of a pseudogap is unexceptionally observed regardless of the photon energy and samples (Extended Data Fig. 4).

## Theoretical model of liquid metals

The backward-bending band dispersion and gaplike features can be naturally explained by the characteristic feature in the band structure predicted in the theory of liquid metals[1-12]. We employ a single-site (structure-independent) model[12] for multiple scattering (Methods), in which each ion is assumed to be a spherical step potential $U_s$ whose depth is $V_0$ and radius is $r_s$ (Fig. 3a). This may be a crude approximation, but as will be shown below, even this simple model is able to capture the essential feature in the band structure of liquid metals. For the partial wave of $l \neq 0$[9], the sum of ionic $U_s$ and centrifugal $U_l$ forms a potential well around each ion. Electron waves scattered by this effective potential acquire a phase shift $\delta_l$ as a function of $k$, such as that in Fig. 3b. If $\delta_l$ rises through $\pi/2$ in the $l^\text{th}$ partial wave, the magnitude of scattering amplitude $f_l$ is peaked at $k_r$, which is the characteristic feature of resonance[9] (Fig. 3c). The phase difference between incoming and outgoing waves outside of $r_s$ is $2\delta_l$ that is $\pi$ at $k_r$. Their destructive interference spatially localizes the electron waves in the potential well around each ion, which is the formation of QBS (Fig. 3d).

The variation of complex $kf_l$ with $\delta_l$ is restricted by the unitary relation[29] that can be readily understood by drawing the unitary circle in Fig. 3e. This explains that Re($kf_l$) and Im($kf_l$) take the simple form of $\sin(2\delta_l)$ and $\sin^2(\delta_l)$, respectively (Fig. 3f). In the theory of liquid metals[1-9] (Methods), a complex $\Delta k$ of electron waves acquired by the effect of multiple scattering can be approximately written in terms of the predominant $f_l$ of the partial wave at resonance as:

$$\Delta k \sim \frac{2\pi n_d f_l}{k} = \frac{2\pi n_d}{k^2} \sin \delta_l \, e^{i\delta_l}.$$

It follows that Re($\Delta k$) and Im($\Delta k$) should be directly proportional to $\sin(2\delta_l)/k^2$ and $\sin^2(\delta_l)/k^2$, respectively. As depicted in Fig. 3g, the typical quadratic band dispersion of free electrons (dotted line) is distorted by Re($\Delta k$) in the sinusoidal form (bold line). The Fourier transform of electrons with complex $\Delta k$ is peaked at $k$ + Re($\Delta k$), but spreads over the range of ±Im($\Delta k$) (grey area in Fig. 3g). This spread of $k$ corresponds to the spatial localization of electrons or QBS in the potential well around each ion (Fig. 3a,d). This is accompanied by changes in the density of states[5,9-12] that is roughly proportional to the energy derivative of Re($\Delta k$) (Fig. 3h). The negative slope of band dispersion near $E_r$ (shown in red) forms a local minimum in the density of states, which is the pseudogap[13,14]. If a residual density of states in the pseudogap is below about 1/3 of that of free electrons[14], electronic states in the pseudogap are localized as shown in Fig. 3d, so that conductivity vanishes in the sense of Anderson localization[15].

**Partial-wave analysis and simulations**

The resonance scattering, which is characterized by $\delta_l$ that rises through $\pi/2$ (Fig. 3b), occurs in one of the partial waves ($l \neq 0$) for a given $k_r$ (Extended Data Fig. 5)[9]. Which partial wave is at resonance depends on the depth of scattering potential $V_0$ (Fig. 3a). As shown in Fig. 4a, there are three distinct $V_0$ ranges to have the partial wave of $l = 1$ ($p$-wave), $l = 2$ ($d$-wave), and $l = 3$ ($f$-wave) at resonance, respectively. The variation of $\delta_l$ with $k$ is calculated for 3 different $V_0$'s to have $p$-wave, $d$-wave, and $f$-wave resonance at the same $k_r$ (dotted line in Fig. 4a), as compared in Fig. 4b. The rise of $\delta_l$ at $d$-wave resonance is steeper (narrower in $k$) than that at $p$-wave resonance. The corresponding density of states is calculated based on thin-slab approximations (Methods)[12], which is shown in Fig. 4c. Indeed, $p$-wave pseudogap is greater in magnitude than $d$-wave pseudogap. Another remarkable difference in Fig. 4b is that $\delta_l$ at $p$-wave resonance rises through $\pi/2$ but turns back without reaching close to $\pi$ (incomplete resonance, Extended Data Fig. 6). This makes $p$-wave pseudogap less clear with a residual density of states, as compared to $d$-wave pseudogap (Fig. 4c).

The band structure renormalized by resonance scattering as well as the calculated density of states are taken to simulate a series of ARPES spectra (Methods), as displayed in Fig. 4e-j. The simulated APRES spectra for $p$-wave resonance in Fig. 4e-g reproduce the backward-bending dispersion of Na-doped black phosphorus with a diminishing intensity towards $E_F$ (Fig. 2d-f). This could have been observed owing to the finite density of states in $p$-wave pseudogap (red curve in Fig. 4c). On the other hand, the sharp dip of $d$-wave pseudogap (yellow curve in Fig. 4c) renders the simulated ARPES spectra in Fig. 4h-j to exhibit a clear gaplike feature with little spectral weight on the backward-bending part of band dispersion, as observed for K, Rb, and Cs in Fig. 2g-o (Extended Data Fig. 7 for a fuller set of simulations). Therefore, our spectral simulations (Fig. 4e-j) not only collectively reproduce key aspects of experimental observations (Fig. 2d-o), but also naturally explain the difference between the Na case and K, Rb, Cs cases by the orbital character of pseudogap, $p$-wave or $d$-wave. This is further supported by the $\delta_l$ of partial waves calculated with the realistic screened scattering potential of alkali ions (Extended Data Fig. 8)[30], where K, Rb, and Cs show $d$-wave resonance, whereas Na shows $p$-wave resonance, as exactly observed in our experimental results.

For the quantitative analysis on the magnitude of pseudogap, the ARPES spectral weight, which is proportional to the density of states, is plotted as a function of energy in Fig. 4d for Na and K (Methods). They commonly show the diminishing spectral weight towards $E_F$ over different energy scales. The magnitude of pseudogap is defined by the energy at which the spectral weight drops by half relative to $E_F$. A fit to this pseudogap with that in the calculated density of states (black curves in Fig. 4d, see Methods) yields 235±53 meV for Na, 65±23 meV for K, 65±17 meV for Rb, and 33±12 meV for Cs. They are in the range of *p*-wave (200–250 meV) and *d*-wave (30–80 meV) pseudogaps at their corresponding $k_r$, respectively (Fig. 5a).

## Doping dependence and phase diagram

The magnitude of pseudogap is plotted in Fig. 5b as a function of the dopant density $n_d$ for different kinds of alkali metal (Extended Data Fig. 9 for a doping series of raw ARPES data). In the wide range of $n_d$ shown in grey, the pseudogap was persistently observed at $E_F$. Since $k_r$ is inversely proportional to the mean interatomic distance of dopants, the inverse square of which corresponds to $n_d$, that is, $n_d \sim k_r^2$. For an insulator doped by monovalent dopants, $n_d$ equals to $n_e$ that is proportional to the square of $k_F$, that is, $n_e \sim k_F^2$. Therefore, the rate of increase in $k_r$ as a function of $n_d$ should be the same as that in $k_F$, which means that $k_r$ follows $k_F$ ($E_r$ is always located at $E_F$) as $n_d$ increases. Once this subtle balance between the rate of increase in $k_F$ and $k_r$ is broken by the C2 band crossing $E_F$, $n_d$ is no longer equivalent to the $n_e$ of C1, and the magnitude of a pseudogap reduces as observed in Fig. 5b. This can also naturally explain the absence of pseudogap in C2 by the deviation of $k_F$ from $k_r$ (0.2-0.3 Å$^{-1}$), which is clearly greater than half the $\Delta k$ of pseudogap (at most 0.08 Å$^{-1}$ for Na).

It is also remarkable that the pseudogap observed in the wide range of $n_d$ is nearly isotropic in magnitude. The pseudogap can be isotropic if the spatial distribution of dopants is made to locate a set of $k_r$ (shown by the blue oval in Fig. 1c) along the Fermi surface of C1 (Fig. 1d), which is supported by the radial and anisotropic structure factor reported in the STM study[25] at $n_d = 1.8 \times 10^{13}$ cm$^{-2}$. This is a natural extension of electronic interactions between dopants mediated by substrates[25,27] with extra energy gains arising from the opening of a pseudogap in the band structure, which is connected to the concept of glassy charge-density waves[31].

## Discussion

It would be difficult to distinguish the ARPES spectra of *d*-wave pseudogap (Fig. 4h-j) from those of actual energy gaps caused by any kind of symmetry breaking in a crystalline solid unless there are other signatures, such as the band folding[32,33] and Bogoliubov bands[34]. Our systematic observation of both *p*-wave and *d*-wave pseudogap allows us to clearly identify their origin as *k* renormalizations induced by resonance scattering, which were predicted in theoretical models for the band structure of liquid metals. This renormalized band structure is a common characteristic regardless of dimensionality, but it may affect the transport of QBS as in Anderson localization[15], depending on how well confined by the potential of ions. It should be noted again that the backward-bending band dispersion of Na-doped black phosphorus in Fig. 2f even reaches close to $k = 0$ ($|\Delta k|$ is comparable to the magnitude of $k$), as predicted in the self-consistent model proposed by Anderson and McMillan[5]. A potential pairing instability in this band structure might be an interesting topic for future works.

Our result may provide a clue to the puzzling spectrum of other crystalline insulators doped by disordered dopants. One such example is the waterfall dispersion widely observed and hotly debated in the study of cuprates[16-18]. Extended Data Fig. 10a shows ARPES data taken from $Bi_{1.5}Pb_{0.5}Sr_2CaCu_2O_{8+\delta}$ ($T_c$ = 88 K), where there is a clear signature of waterfall dispersion. $n_d$ is estimated from $T_c$[35], pseudogap (32 meV)[36], and Luttinger count as $1.2-1.4 \times 10^{14}$ cm$^{-2}$, corresponding to $k_r = 0.29 \pm 0.01$ Å$^{-1}$. Our simulation for resonance scattering at $k_r = 0.29$ Å$^{-1}$ in Extended Data Fig. 10b (Methods) reproduces the waterfall dispersion remarkably well. As such, the physics revealed here may play a role in the puzzling spectrum of cuprates[16-18] and other crystalline insulators doped by disordered dopants[19,20].

**Publisher's note:** Springer Nature remains neutral with regard to jurisdictional claims in published maps and institutional affiliations.

**Acknowledgements** We thank S. -S. Baik, Y. -W. Son, and E. -G. Moon for helpful discussions, D. Song and C. Kim for providing samples. This work was supported by National Research Foundation (NRF) of Korea (Grants No. NRF-2020R1A2C2102469, NRF-2017R1A5A1014862, NRF-2020K1A3A7A09080364), and the Yonsei Signature Research Cluster Program of 2021 (2021-22-0004). This research used resources of the Advanced Light Source, which is a DOE Office of Science User Facility under Contract No. DE-AC02-05CH11231.

**Author contributions** S.H.R., M.H., and D.Y.P. performed ARPES experiments with help from C.J., E.R., and A.B. S.H.R. and M.H. performed model calculations and simulations with help from D.Y.P. K.S.K. conceived and directed the project. S.H.R., M.H., D.Y.P., and K.S.K wrote the manuscript with contributions from all other co-authors.

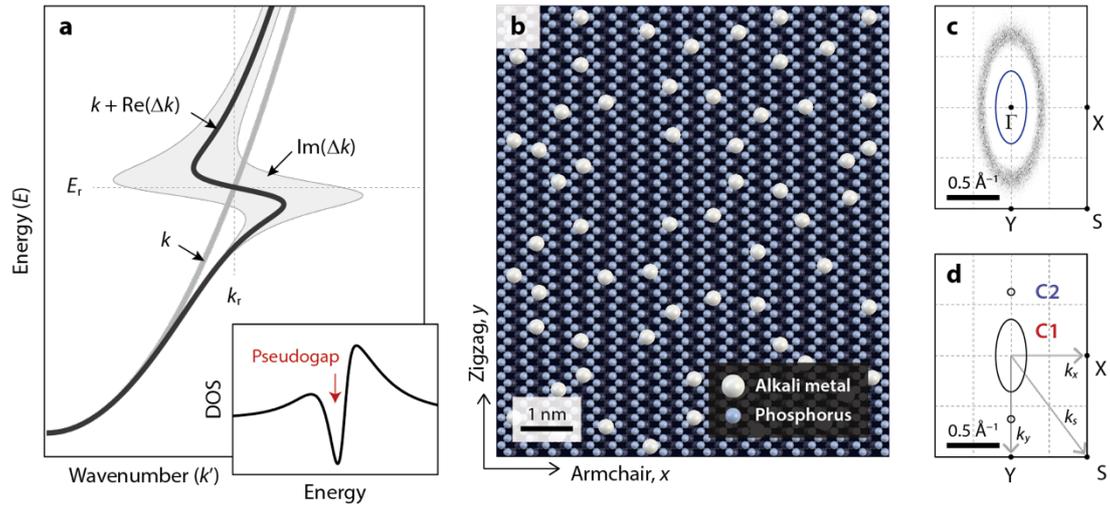

**Figure 1 | Band structure of liquid metals and experimental systems. a,** Electronic band structure predicted in the theory of liquid metals[1-12]. The black bold line shows a dispersion relation of energy ($E$) versus $k'$ renormalized by $\mathrm{Re}(\Delta k)$ from the quadratic band structure of free electrons ($E \sim k^2$) shown by the grey bold line. The grey region represents the spread of $k$ over the range of $\pm\mathrm{Im}(\Delta k)$ with respect to $E_r$ and $k_r$. Inset shows the corresponding density of states as a result of $k$ renormalizations. **b,** Schematic of disordered alkali metals on black phosphorus simulated based on scanning tunnelling microscopy (STM) results[25] (Methods). **c,** Fast Fourier-transform (FFT) image taken from the distribution of dopants in **b** to show their reciprocal lattice. The blue oval is a set of half the reciprocal lattice vectors in every in-plane $k$ direction, which corresponds to $k_r$ in the model of liquid metals. **d,** Constant-energy contours at $E_F$ expected for black phosphorus whose surface is doped at $n_e = 1.0 \times 10^{14}$ cm$^{-2}$ and plotted over the surface Brillouin zone. The grey arrows indicate 3 high-symmetry $k$ cuts along which a series of ARPES data shown in Fig. 2 is taken.

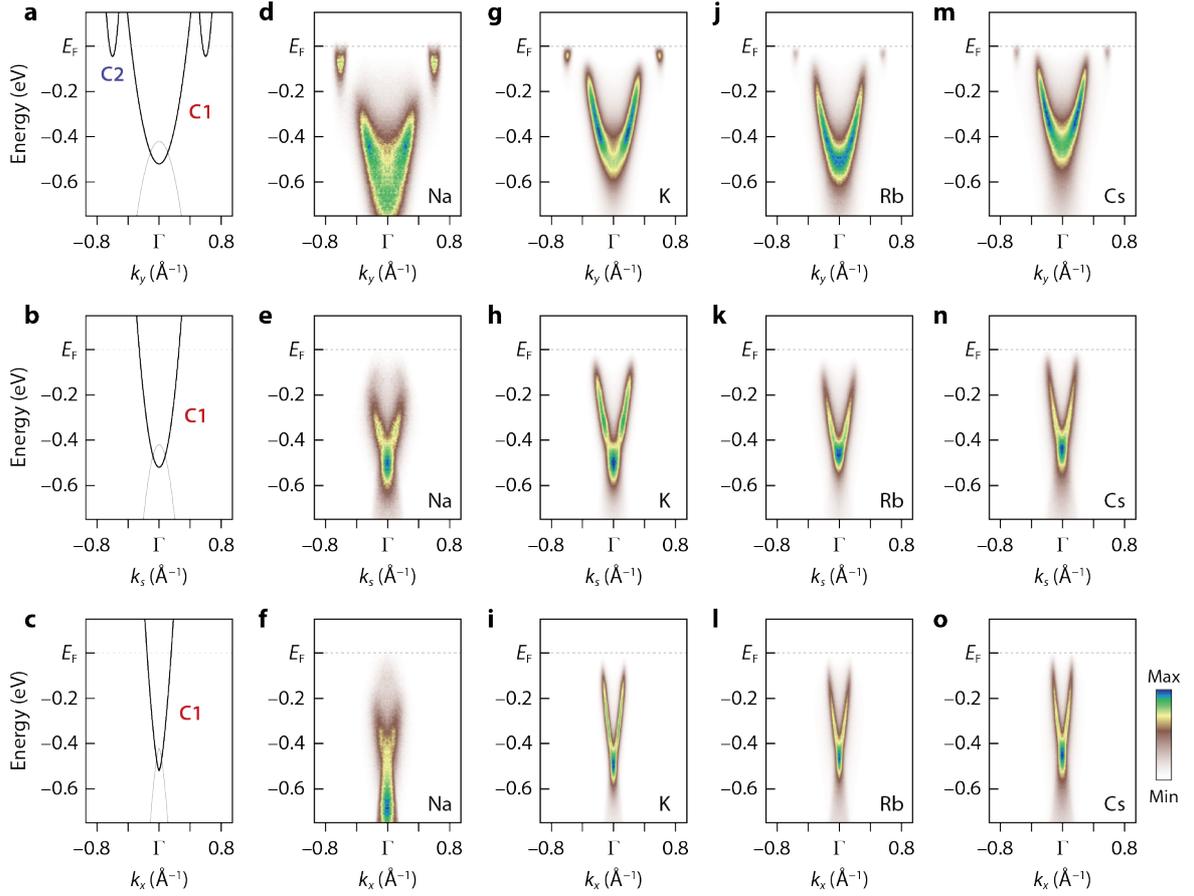

**Figure 2 | Electronic structure of black phosphorus doped by disordered alkali metals.**
**a-c,** Conduction bands of black phosphorus doped to locate the bottom of C2 bands at $E_F$ ($n_e = 1.0 \times 10^{14}$ cm$^{-2}$), shown along $k_y$ (**a**), $k_s$ (**b**), and $k_x$ (**c**) indicated by grey arrows in Fig. 1d. **d-o,** Corresponding experimental band structure of bulk black phosphorus whose surface is doped by different kinds of alkali metal, Na (**d-f**), K (**g-i**), Rb (**j-l**), and Cs (**m-o**), as marked at the bottom right of each panel. This series of data is measured by ARPES at 15–35 K with the photon energy of 104 eV along $k_y$ (top row), $k_s$ (middle row), and $k_x$ (bottom row). The grey lines in **a-c** are the valence band whose spectral intensity is fully suppressed in ARPES data along $k_y$[37]. For those taken in $k_s$ and $k_x$, however, this suppression becomes incomplete, making the vertical spread of intensity at below the bottom of C1 (Extended Data Fig. 2).

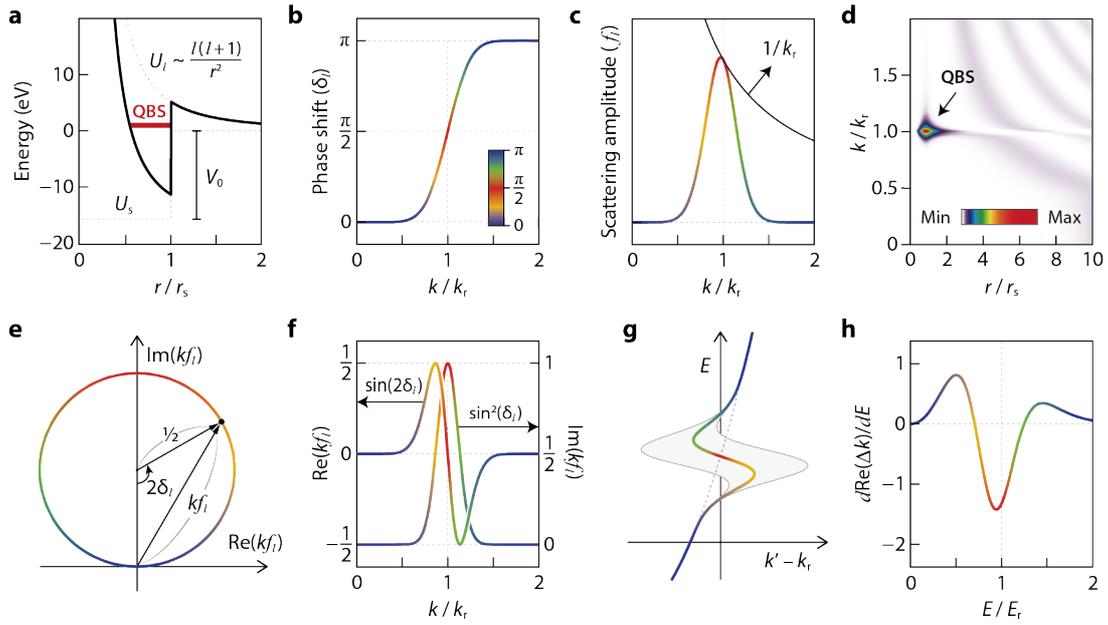

**Figure 3 | Resonance scattering in liquid metals and band renormalizations. a,** Effective scattering potential (bold line) composed of the ionic potential (dotted line marked by $U_s$) and the centrifugal potential (dotted line marked by $U_l$). $U_s$ is assumed as a step potential whose depth is $V_0$ and width is $r_s$. **b,** Phase shift $\delta_l$ simulated in the form of sigmoid that rises from 0 to $\pi$ through $\pi/2$ at $k_r$. **c,** Magnitude of the scattering amplitude $f_l$ calculated from $\delta_l$ in **b** by $\sin(\delta_l)/k$. The black line is a trace of $f_l$ at resonance with $k_r$. **d,** $k$ series of a probability density calculated for spherical waves in the effective potential in **a** (Methods). **e,** Argand diagram of $kf_l$ with the argument of $2\delta_l$. **f,** Re($kf_l$) and Im($kf_l$) calculated from $\delta_l$ in **b** in the form of $\sin(2\delta_l)$ and $\sin^2(\delta_l)$, respectively. The relation of Re($kf_l$) and Im($kf_l$) can be understood by the unitary circle with a radius of 1/2 in **e**. **g,** Band renormalizations in liquid metals: The quadratic band structure of free electrons (dotted line) is renormalized by Re($\Delta k$) (bold line) with a spread of $k$ over the range of $\pm$Im($\Delta k$) (grey area). **h,** Energy derivative of Re($\Delta k$) shown with respect to $E_r$. $\delta_l$ is colour-coded as indicated at the bottom right of **b**, which is used to represent the corresponding value of $\delta_l$ in **c** and **e-h**.

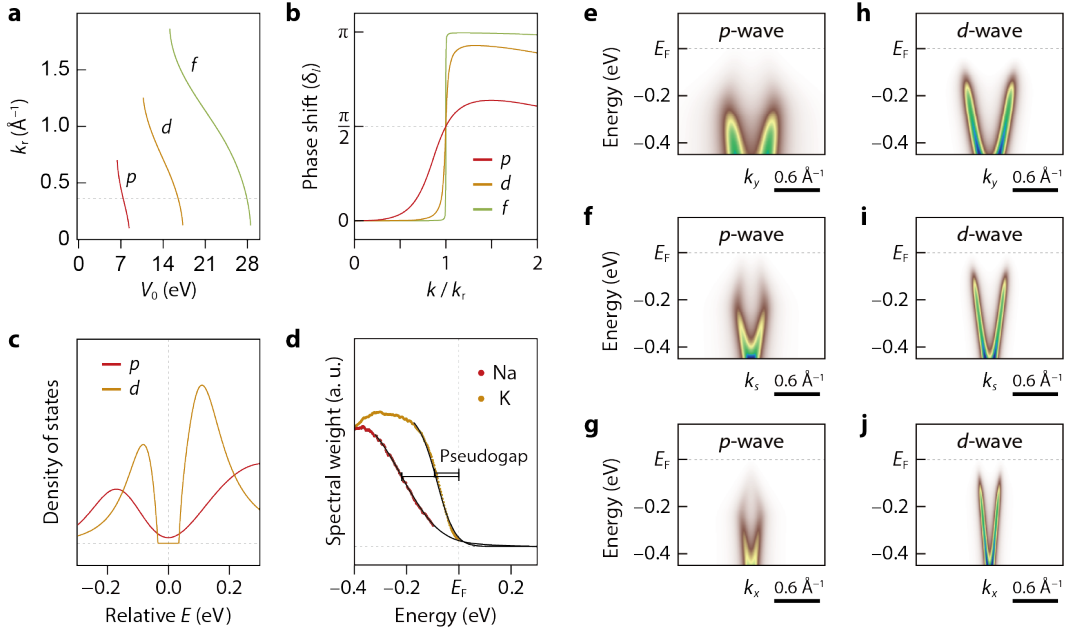

**Figure 4 | Density of states in liquid metals and spectral simulations. a,** Potential depth ($V_0$ in Fig. 3a) to have *p*-wave, *d*-wave, and *f*-wave resonance in the model of liquid metals (Methods). **b,** Comparison of $\delta_l$ for *p*-wave ($l = 1$), *d*-wave ($l = 2$), and *f*-wave ($l = 3$) resonance calculated for 3 different $V_0$'s to have the same $k_r$ indicated by the dotted line in **a**. **c,** Density of states calculated for *p*-wave and *d*-wave resonance (Methods) and plotted in $E$ relative to the local minimum (pseudogap). **d,** Experimental spectral weight obtained by integrating the spectral weight of C1 over the first Brillouin zone. The black curves overlaid are a fit to the pseudogap in the spectral weight with that in the calculated density of states (Methods). The scale bars are the magnitude of pseudogap defined by the energy at which the spectral weight drops by half relative to $E_F$. **e-j,** Spectral simulation of *p*-wave (**e-g**) and *d*-wave (**h-j**) resonance shown along $k_y$ (top row), $k_s$ (middle row), and $k_x$ (bottom row).

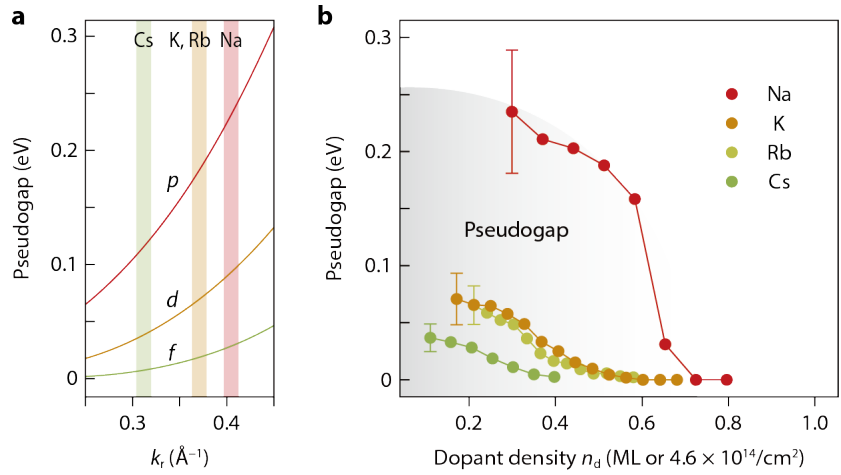

**Figure 5 | Magnitude of the pseudogap and phase diagram. a,** Magnitude of pseudogap calculated for *p*-wave, *d*-wave, and *f*-wave resonance with $k_r$ in Fig. 4a. The vertical stripes marked by Na, K, Rb, and Cs are their respective range of $k_r$, in which a maximal pseudogap is observed in experiments. **b,** Magnitude of pseudogap estimated from ARPES data of black phosphorus doped by Na, K, Rb, and Cs (Extended Data Fig. 9), which is plotted as a function of $n_d$ over the range of 0–1 ML. The grey area represents the pseudogap phase.

## Methods

**ARPES experiments and analysis.** We conducted ARPES measurements at the Beamline 7.0.2 (MAESTRO), the Advanced Light Source. The microARPES end-station is equipped with a hemispherical spectrometer. We collected the data at the sample temperature of 15–35 K with the photon energy of 104–125 eV and linear-horizontal polarization. At these settings, the energy and $k$ resolutions are better than 20 meV and 0.01 Å$^{-1}$, respectively. We carefully aligned the in-plane orientation of black phosphorus along the $k_y$, $k_s$, or $k_x$ direction with respect to the analyser slit oriented in the direction perpendicular to the scattering plane for symmetry. Then, we repeated the cycle of *in situ* deposition and data acquisition to take a doping series of ARPES data shown in Fig. 2d-o and Extended Data Fig. 9. After converting the emission angle to $k$, a series of $k$ distribution curves at a constant energy was fit with the standard Lorentzian function. The spectral weight of C1 integrated in the first Brillouin zone is plotted as a function of energy in Fig. 4d. The energy at which the spectral weight drops by half relative to $E_F$ is taken as the magnitude of pseudogap in Fig 5b. The ARPES data for cuprates (Extended Data Fig. 10a) is taken at Beamline I05, the Diamond Light Source with the sample temperature of 5 K and the photon energy of 20–140 eV.

**Sample preparation and surface doping.** The single-crystal samples of black phosphorus (99.995%, HQ graphene) were glued on sample holders by conductive epoxy. The sample holders were transferred in the ARPES chamber through a semi-automated sample highway system. The black phosphorus samples were cleaved in the ultrahigh vacuum chamber with the base pressure better than $5 \times 10^{-11}$ torr. We scanned over the surface with the photon beam of 80 μm in diameter to carefully optimize the sharpness of ARPES spectra (Fig. 2d-o). The *in situ* deposition of alkali metals (Na, K, Rb, and Cs) was carried out using commercial dispensers (SAES) at the adsorption rate of over $3.6 \times 10^{11}$ cm$^{-2}$/s. $n_d$ is assumed equivalent to $n_e$ estimated from the density of surface unit cells ($6.9 \times 10^{14}$ cm$^{-2}$) multiplied by the ratio of area enclosed by constant-energy contours at $E_F$ (Fig. 1d) to that of the surface Brillouin zone. We extrapolate $k_F$ in $x$ ($k_{F,x}$) and $k_F$ in $y$ ($k_{F,y}$) from the band dispersion of C1, and estimate the area of Fermi contours as $\pi k_{F,x} k_{F,y}$. The unit of ML in Fig. 5b is defined as $4.6 \times 10^{14}$ cm$^{-2}$, which is the density of closely packed K atoms in two dimensions. The single-crystal samples of Bi$_{1.5}$Pb$_{0.5}$Sr$_2$CaCu$_2$O$_{8+\delta}$ overdoped at $T_c = 88$ K was used for Extended Data Fig. 10a.

**Theoretical model for liquid metals.** We employ the theory of multiple scattering[1-12] by each ion assumed as a spherical step potential whose depth is $V_0$ and radius is $r_s$ (Fig. 3a). This boundary-condition problem is solved based on spherical wavefunctions and partial-wave expansion to obtain the phase shift of partial waves $\delta_l$ (Fig. 3b). Then, the scattering amplitude of partial waves $f_l$ can be simply written in terms of $\delta_l$ as:

$$f_l = \frac{\sin \delta_l \, e^{i\delta_l}}{k}.$$

Re($kf_l$) and Im($kf_l$) take the simple mathematical form of sin($2\delta_l$) and $\sin^2(\delta_l)$ (Fig. 3f), and their relation can be straightforwardly shown by the unitary circle in the complex plane (Fig. 3e)[29]. The magnitude of $f_l$ in Fig. 3c is given by sin($\delta_l$)/$k$ with its maximum possible value of $1/k_r$ at resonance. By taking the square of spherical wavefunctions at a constant $k$, the probability density is calculated and plotted as a function of $r$ and $k$ in Fig. 3d after the normalization of the first peaks outside of $r_s$. In the thin-slab approximation ($k \gg \Delta k$, forward scattering)[9,12], $\Delta k$ can be written in terms of the predominant $f_l$ of the partial wave at resonance as:

$$\Delta k \sim \frac{2\pi n_d f_l}{k}.$$

It follows that Re($\Delta k$) and Im($\Delta k$) are directly proportional to sin($2\delta_l$)/$k^2$ and $\sin^2(\delta_l)/k^2$ with respect to $k_r$, respectively. The quadratic band structure of $E \sim k^2$ is renormalized by Re($\Delta k$) as $k' = k +$ Re($\Delta k$) with the spread of $k'$ over the range of $\pm$Im($\Delta k$), which is the band structure of liquid metals depicted in Fig. 3g.

The resonance scattering, which is characterized by $\delta_l$ that passes through $\pi/2$, occurs in one of the partial waves for a given $k_r$. Which partial wave is at resonance scattering depends on the depth of scattering potential $V_0$ in Fig. 3a. For a constant $V_0$ (set to 16.4 eV for $d$-wave resonance), a partial-wave series of $\delta_l$ and $\Delta k$ is calculated as described above and displayed in Extended Data Fig. 5. We systematically calculated $p$-wave, $d$-wave, and $f$-wave resonance as a function of $V_0$ with the constant $r_s$ (set to 2.1 Å for liquid Na[38]), as shown in Fig. 4a. The representative set of $\delta_l$ at $p$-wave, $d$-wave, and $f$-wave resonance in Fig. 4b is calculated for $V_0 = 7.4$ eV, 16.4 eV, and 27.8 eV to have the same $k_r$ of 0.36 Å$^{-1}$ (Extended Data Fig. 6 for the fuller set of $U_l$, $\delta_l$, and $\Delta k$). The density of states in Fig. 4c is calculated using Equation (3.13) in Ref. 12 with the input parameters of $V_0 = 7.15$ eV for $p$-wave resonance and $V_0 = 16.26$ eV for $d$-wave resonance at constant $r_s = 2.1$ Å. The half width at half maximum of Im($\Delta k$), which corresponds to half the dip width in the energy derivative of Re($\Delta k$) in Fig. 3h, is taken as the magnitude of pseudogap in Fig. 5a.

**Spectral simulations.** The band structure of black phosphorus doped by disordered metals is simulated based on the model of $k$ renormalizations. The $k$ distribution of ARPES intensity is modelled in the form of Lorentzian, the integration of which over $k$ is a constant function with energy. The energy distribution of spectral weight, which is proportional to the density of states, is given by $n_E$ that accounts for the diminishing spectral weight towards $E_F$ by the formation of pseudogap as:

$$I_E(k') \sim \frac{[\text{Im}(\Delta k)_E + \eta]}{[k' - k_E - \text{Re}(\Delta k)_E]^2 + [\text{Im}(\Delta k)_E + \eta]^2} \cdot f_{\text{FD}} \cdot n_E$$

where $k_E$ is the dispersion of non-interacting bands in a form of $E \sim k^p$ with $k_F$ and the bottom energy taken from ARPES data. $p$ is set to 1.8–2.0 (quadratic) in $y$ and 1.2–1.4 (nearly linear) in *x* to reflect the well-known armchair-zigzag anisotropy of black phosphorus[23]. Re($\Delta k$) and Im($\Delta k$) are taken in a form of $\sin(2\delta_l)/k^2$ and $\sin^2(\delta_l)/k^2$, respectively, with respect to $k_r$ located near $k_F$. The magnitude of Re($\Delta k$) and Im($\Delta k$) is scaled to fit the spectral peak position and width of ARPES spectra. $\eta$ is the offset *k* broadening in the range of 0.03–0.14 Å$^{-1}$, and $f_{FD}$ is the Fermi-Dirac distribution function.

$n_E$ is the density of states calculated using Equation (3.13) in Ref. 12 and convoluted by the error function to account for finite experimental broadening. The input parameters are $V_0$ and $r_s$, the latter of which is set to 2.1 Å. $V_0$ is optimized to reproduce the magnitude of a pseudogap in the spectral weight obtained by integrating the ARPES intensity of C1 over all of *k* space in the first Brillouin zone in Fig. 4d, which yields 7.15 eV for Na and 16.26 eV for K at experimental broadening of 0.06–0.14 eV. The simulated spectra are shown in Fig. 4e-j, and directly compared to a fuller set of data in Extended Data Fig. 7.

As for Extended Data Fig. 10b, $k_E$ is assumed to be the quadratic band dispersion shown by the grey curve. The $n_d$ of our samples is estimated from $T_c$ (88 K)[35], pseudogap (32 meV)[36], and the Luttinger count to be 1.2–1.4 × 10$^{14}$ cm$^{-2}$, corresponding to the $k_r$ of 0.29 ± 0.01 Å$^{-1}$. Re($\Delta k$) and Im($\Delta k$) are calculated based on the model of liquid metals with input parameters of $V_0$ = 9 eV and $r_s$ = 2 Å for the *p*-wave resonance at $k_r$ = 0.29 Å$^{-1}$. $\eta$ is set to be 0.01 Å$^{-1}$ at $E_F$, from which it monotonically increases with the binding energy up to 0.08 Å$^{-1}$.

**Structural simulations.** We consider 7200 dopants distributed in the area of 150 × 150 nm$^2$ ($n_d$ = 4.3 × 10$^{13}$ cm$^{-2}$), each of which is constructed by a two-dimensional Gaussian function with the width of 1.5 Å. Considering the Coulomb repulsion between the ionized dopants, we employ the hard-sphere model in which the position of atoms is randomized under the condition that the interatomic distance cannot be smaller than the radius of hard spheres *d* (note that *d* is scaled to the mean interatomic distance, which is 1.2 nm for $n_d$ = 4.3 × 10$^{13}$ cm$^{-2}$). In the FFT image of simulated structures, one can always find a radial structure factor, as shown in Extended Data Fig. 1g. This radial shape itself in the structure factor reflects the presence of the mean interatomic distance of dopants (or only short-range order) generic to ionized dopants randomly distributed under the repulsive interaction. To account for the isotropic magnitude of pseudogap over the anisotropic Fermi surface of C1, the anisotropy of the structure factor is additionally considered based on the recent STM study[25], where the radial and anisotropic structure factor was clearly found at the density of 1.8 × 10$^{13}$ cm$^{-2}$. The anisotropy factor of 0.42 taken from the Fermi surface in Fig. 1d is used to simulate the distribution of dopants in Fig. 1b, which shows the radial and anisotropic structure factor in its FFT (Fig. 1c) in excellent agreement with that observed in Ref. 25.

## Data availability

The data that support the findings of this study are available within the paper and from the corresponding author upon reasonable request. Source data are provided with this paper.

**Competing interests** The authors declare no competing interests.

**Additional information**
**Correspondence and requests for materials** should be addressed to K.S.K.
**Peer review information** *Nature* thanks --- for their contribution to the peer review of this work.
**Reprints and permissions information** is available at www.nature.com/reprints.

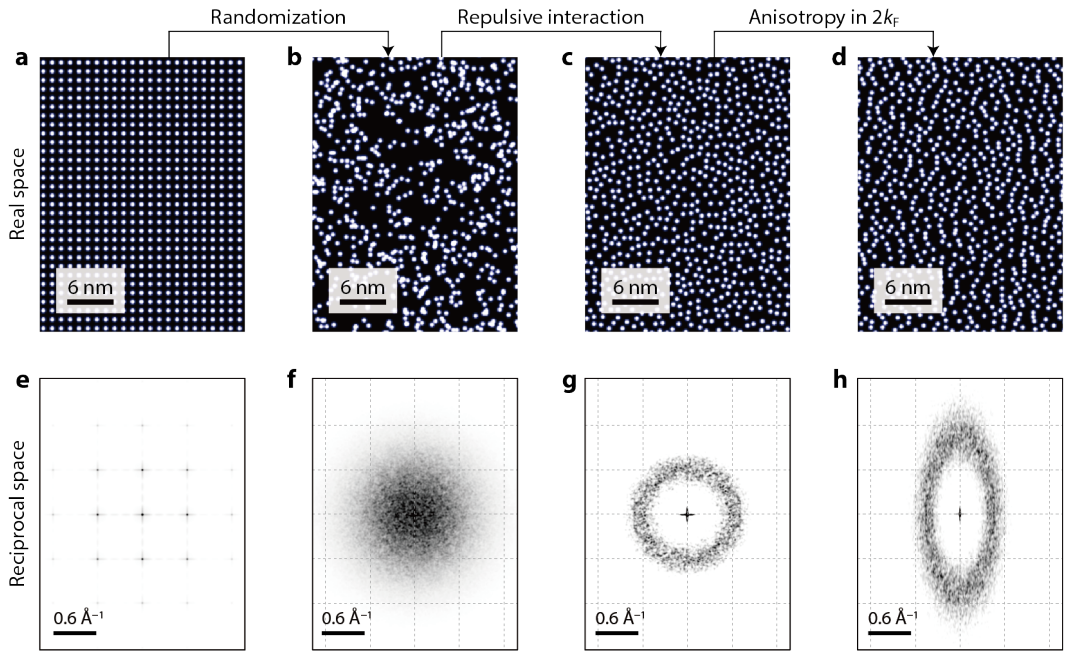

**Extended Data Figure 1 | Key assumptions in structural simulations.** Let us consider the long-range order of dopants whose density is $1 \times 10^{14}$ cm$^{-2}$ (**a**). If the position of dopants is fully randomized (**b**), no structure is expected in its FFT (**f**). Considering that each dopant is ionized by donating its valence electron, there should be the Coulomb repulsion between the ionized dopants, which prevent them to be located close to each other in a certain range corresponding to the radius of hard spheres. If the position of dopants is randomized under this hard-sphere assumption with no further interactions considered (**c**), one can always find the radial structure factor in its FFT (**g**). The radial shape itself in the structure factor reflects the presence of a mean interatomic distance (or short-range order) generic to ionic dopants randomly distributed under the repulsive interaction. This is the structural factor required for resonance effect in the theory of liquid metals[1-12], which is clearly distinguished from both crystalline (**a,e**) and fully random (**b,f**) cases. In case of monovalent dopants on a crystalline insulator, the area of a circle whose radius is half the radial peak is the same as that enclosed by the Fermi surface in any shape due to the charge conservation. To account for the isotropic magnitude of a pseudogap over the anisotropic Fermi surface of C1, the anisotropy of the structure factor is additionally considered based on the recent STM study[25], where the radial and anisotropic structure factor was found at the density of $1.8 \times 10^{13}$ cm$^{-2}$. As the anisotropy factor of 0.42 taken from the anisotropic $k_F$'s in the Fermi surface is applied to the hard-sphere model (**d**), the radial structure factor becomes anisotropic as seen in **h**.

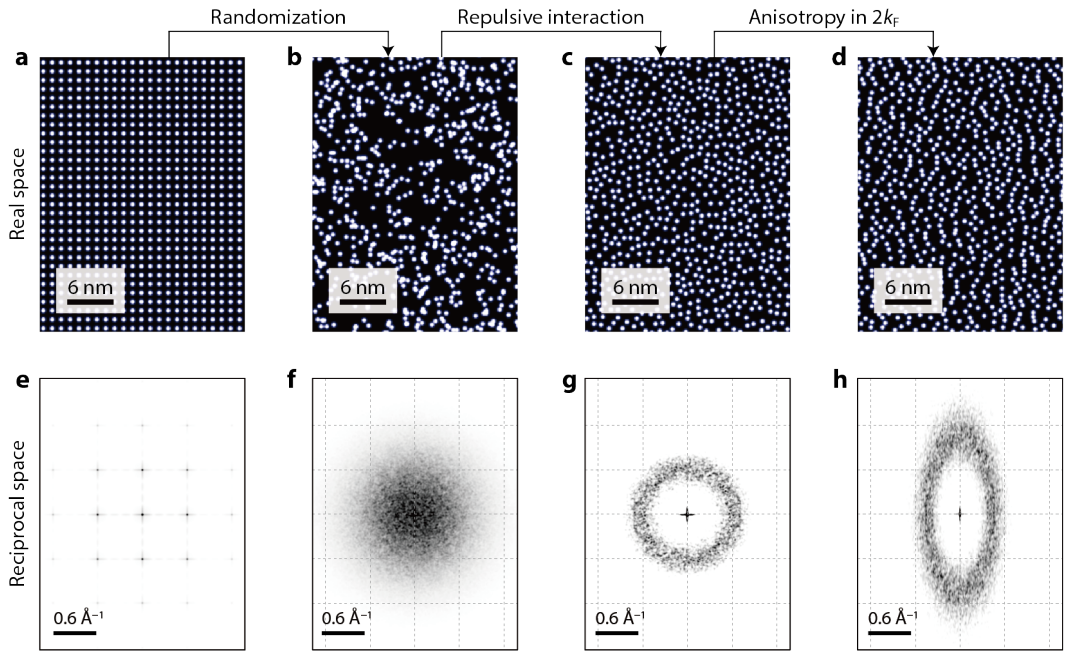

**Extended Data Figure 1 | Key assumptions in structural simulations.** Let us consider the long-range order of dopants whose density is $1 \times 10^{14}$ cm$^{-2}$ (**a**). If the position of dopants is fully randomized (**b**), no structure is expected in its FFT (**f**). Considering that each dopant is ionized by donating its valence electron, there should be the Coulomb repulsion between the ionized dopants, which prevent them to be located close to each other in a certain range corresponding to the radius of hard spheres. If the position of dopants is randomized under this hard-sphere assumption with no further interactions considered (**c**), one can always find the radial structure factor in its FFT (**g**). The radial shape itself in the structure factor reflects the presence of a mean interatomic distance (or short-range order) generic to ionic dopants randomly distributed under the repulsive interaction. This is the structural factor required for resonance effect in the theory of liquid metals[1-12], which is clearly distinguished from both crystalline (**a,e**) and fully random (**b,f**) cases. In case of monovalent dopants on a crystalline insulator, the area of a circle whose radius is half the radial peak is the same as that enclosed by the Fermi surface in any shape due to the charge conservation. To account for the isotropic magnitude of a pseudogap over the anisotropic Fermi surface of C1, the anisotropy of the structure factor is additionally considered based on the recent STM study[25], where the radial and anisotropic structure factor was found at the density of $1.8 \times 10^{13}$ cm$^{-2}$. As the anisotropy factor of 0.42 taken from the anisotropic $k_F$'s in the Fermi surface is applied to the hard-sphere model (**d**), the radial structure factor becomes anisotropic as seen in **h**.

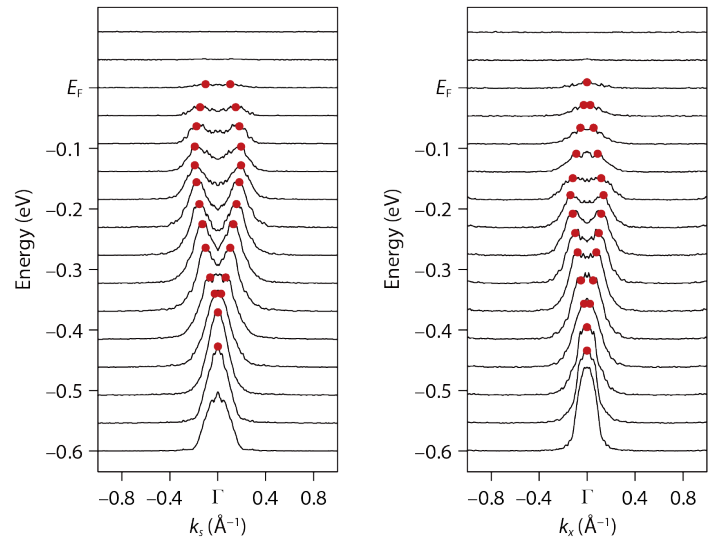

**Extended Data Figure 2 | Momentum distribution curves.** ARPES data of Na-doped black phosphorus taken along $k_s$ (left) and $k_x$ (right) directions (those in Fig. 2e,f). ARPES intensity is plotted as a function of $k$ at the constant energy shown on the vertical axis. Red dots mark the peak position of $k$ distribution curves to show the backward bending band dispersion.

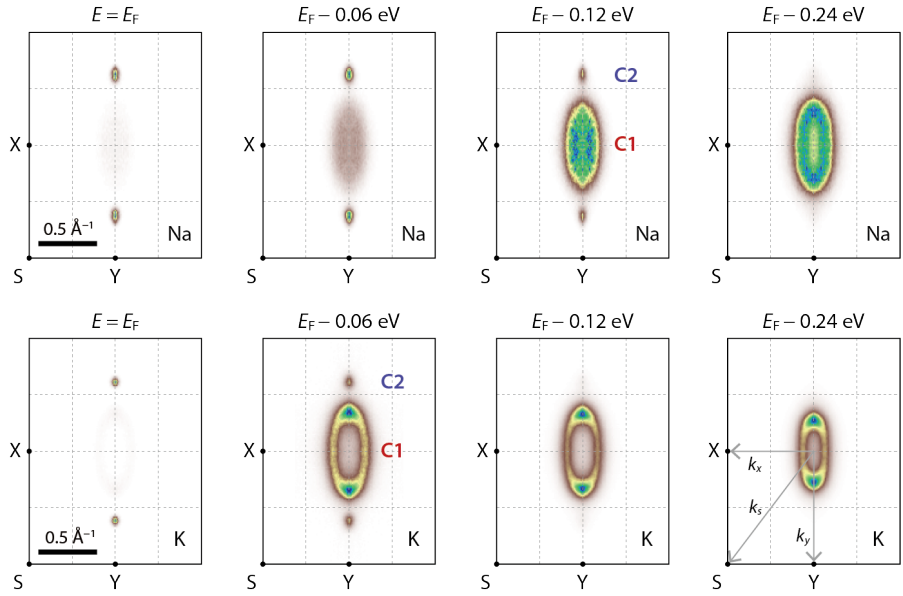

**Extended Data Figure 3 | Constant-energy map of ARPES spectra.** ARPES intensity maps of bulk black phosphorus whose surface is doped by Na at $n_d = 1.7 \times 10^{14}$ cm$^{-2}$ (upper row) and K at $n_d = 1.0 \times 10^{14}$ cm$^{-2}$ (lower row). Data are plotted at the constant energy marked on top of each panel as a function of $k_x$ and $k_y$ in the surface Brillouin zone of black phosphorus. Unlike a pair of C2 bands that crosses $E_F$, the C1 band shows a clear signature of pseudogap with the magnitude of 235 meV for Na and 65 meV for K. This magnitude of pseudogap is found nearly isotropic with in-plane $k$ directions. The size of area enclosed by the C1 contour of Na decreases towards $E_F$, which reflects the backward-bending band dispersion.

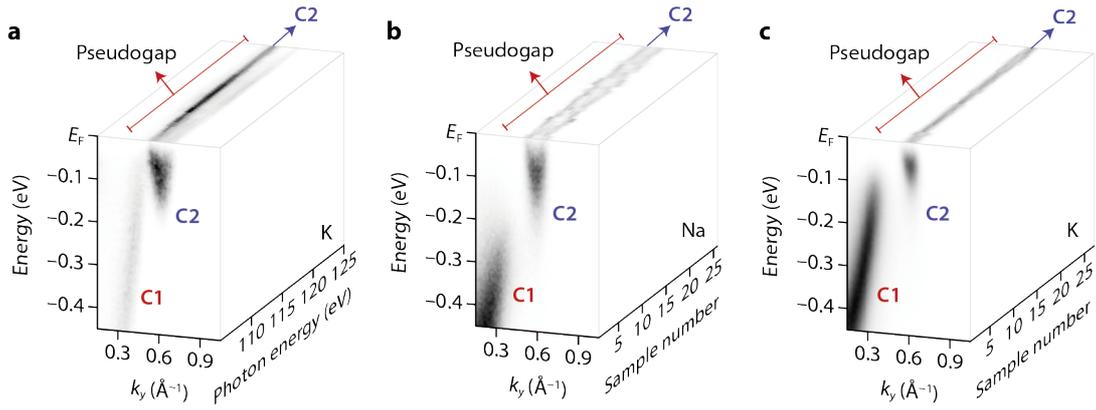

**Extended Data Figure 4 | Dependence of a pseudogap on photon energy and samples.
a,** Photon-energy dependence of ARPES data taken from black phosphorus whose surface is doped by K at $n_d = 1.8 \times 10^{14}$ cm$^{-2}$. The pseudogap of 30 meV at this $n_d$ is robustly observed for the photon energy of 106–125 eV (red arrow) in contrast to the C2 band that crosses $E_F$ (blue arrow). **b,c,** Sample dependence of ARPES data taken from black phosphorus doped by Na at $n_d = 1.7 \times 10^{14}$ cm$^{-2}$ (**b**) and K at $n_d = 1.0 \times 10^{14}$ cm$^{-2}$ (**c**). The pseudogap of 235 meV for Na and 65 meV for K is unexceptionally observed for all of the samples.

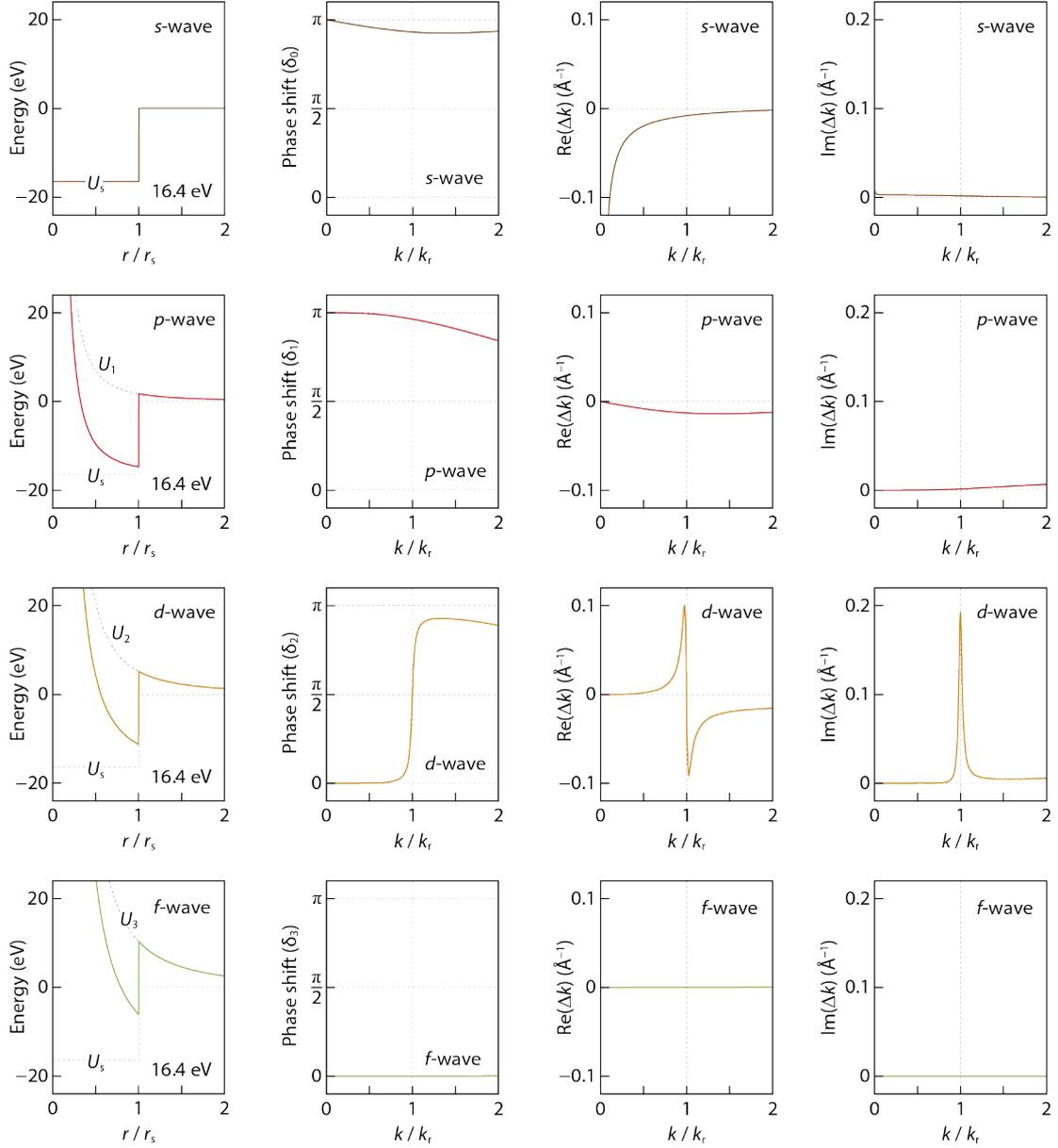

**Extended Data Figure 5 | Model calculation for partial waves at *d*-wave resonance.** A series of $U_l$, $\delta_l$, Re($\Delta k$), and Im($\Delta k$) is calculated as described in Methods for the partial waves from $l = 0$ (top row) to $l = 3$ (bottom row). The depth of potential $V_0$ is set to 16.4 eV, which corresponds to *d*-wave resonance at $k_r = 0.36$ Å$^{-1}$, as shown in Fig. 4a. The *d*-wave resonance can be identified by $\delta_2$ passing through $\pi/2$, which is accompanied by Re($\Delta k$) and Im($\Delta k$) taking the form of $\sin(2\delta_l)/k^2$ and $\sin^2(\delta_l)/k^2$, respectively. For $l = 0$, there is no potential well, in which electron waves can be trapped, and it is impossible to have *s*-wave resonance.

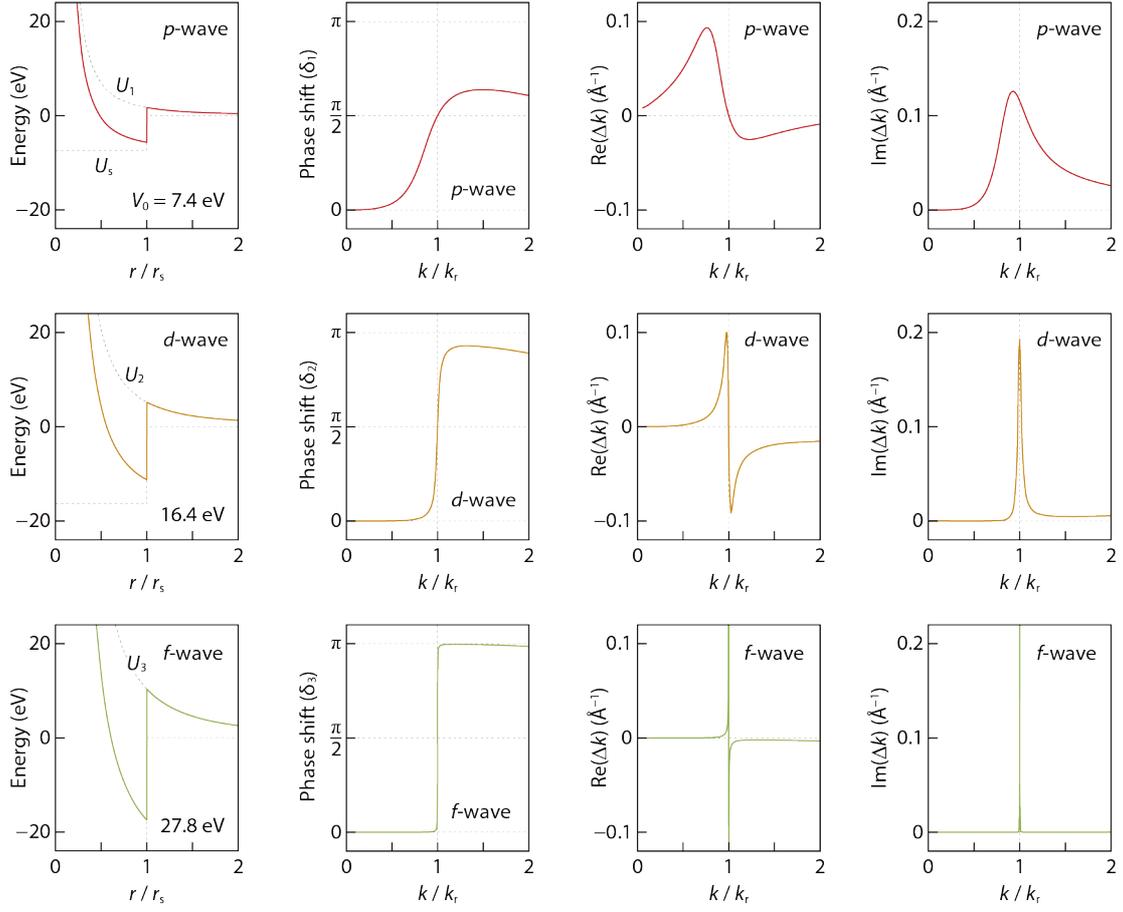

**Extended Data Figure 6 | Comparison of $\Delta k$ at *p*-wave, *d*-wave, and *f*-wave resonance.** A series of $U_l$, $\delta_l$, Re($\Delta k$), and Im($\Delta k$) is calculated for *p*-wave (top row), *d*-wave (middle row), and *f*-wave (bottom row) resonance (Methods). $V_0$ was set to 7.4 eV for *p*-wave, 16.4 eV for *d*-wave, 27.8 eV for *f*-wave resonance to have the same $k_r$ of 0.36 Å$^{-1}$ (dotted line in Fig. 4a). The variation of $\delta_l$, Re($\Delta k$), and Im($\Delta k$) is narrower in $k$ or $E$ width for the higher number of $l$, which accounts for the smaller magnitude of pseudogap in Fig. 5a.

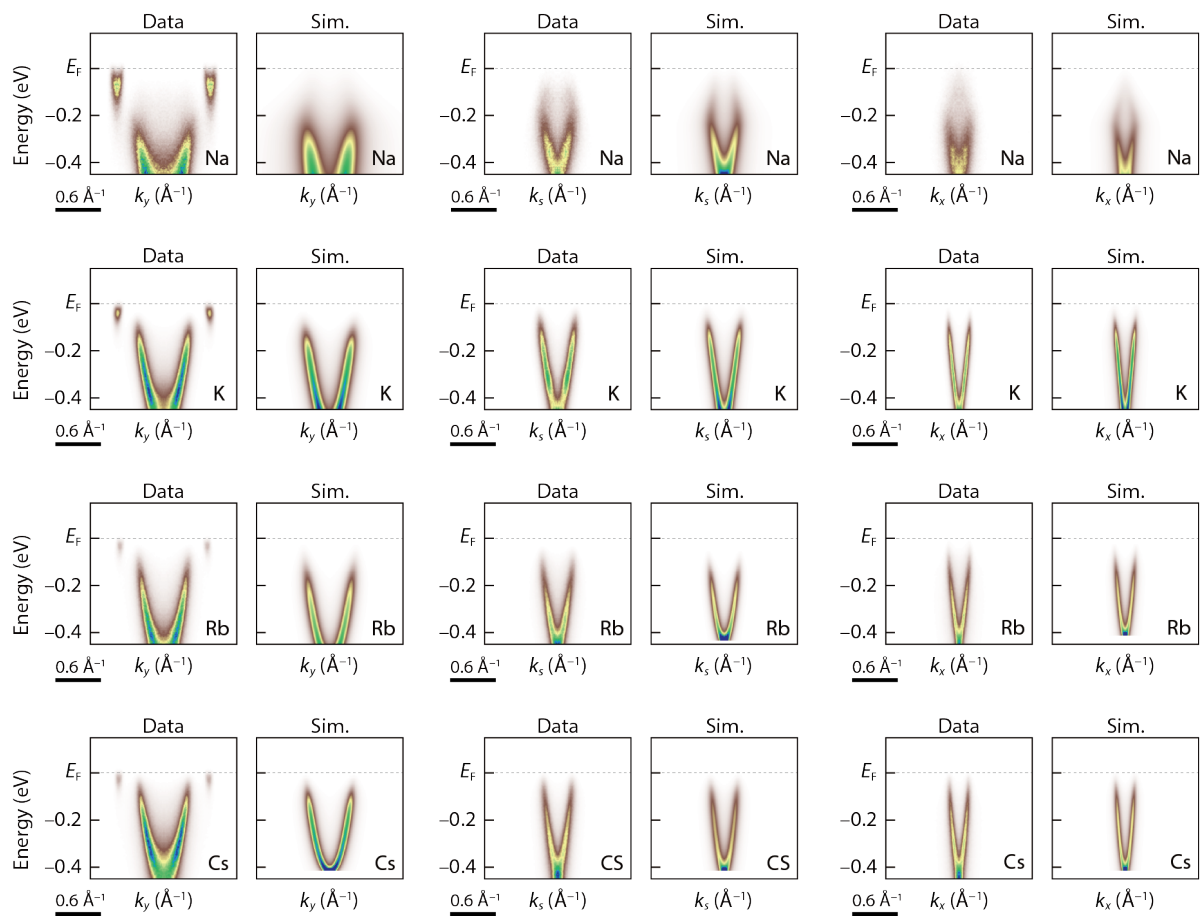

**Extended Data Figure 7 | Spectral simulations compared to experimental ARPES data.** The experimental band structure is taken from black phosphorus whose surface is doped by different kinds of alkali metal (Na, K, Rb, and Cs), as marked at the bottom right of each panel. This series of ARPES spectra was taken along $k_y$ (left column), $k_s$ (middle column), and $k_x$ (right column). Each data is directly compared to corresponding spectral simulations at *p*-wave resonance for Na and *d*-wave resonance for K, Rb, and Cs (Methods). The bottom energy of spectral simulations is limited by that of the non-interacting (bare) band, which is the reason that those for Rb and Cs are slightly cut at the binding energy of 0.42–0.45 eV.

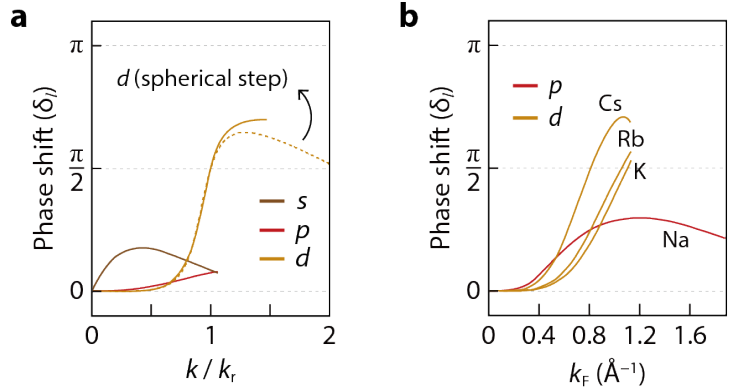

**Extended Data Figure 8 | Partial-wave analysis with the screened scattering potential.**
**a,** Phase shift $\delta_l$ of the partial waves calculated by the screened scattering potential for the liquid phase of Cs ions, which clearly shows *d*-wave resonance[9]. The dotted line is a fit with that calculated by the spherical step potential at $V_0$ = 12.96 eV and $r_s$ = 2.1 Å. **b,** Phase shift $\delta_l$ of partial waves calculated by the screened scattering potential of Na, K, Rb, and Cs ions[30]. K, Rb and Cs ions favour *d*-wave resonance, whereas the Na ions favour *p*-wave resonance, which is exactly as observed in our experiments. Our results reproduce other details as well: (1) $\delta_l$ crosses $\pi/2$ clearly for K, Rb, and Cs ions, while that for Na ions is relatively incomplete. (2) The $\delta_l$'s of K and Rb ions are nearly the same. (3) The abruptness of $\delta_l$, which is related to the magnitude of a pseudogap, increases in the order of Na → K / Rb → Cs.

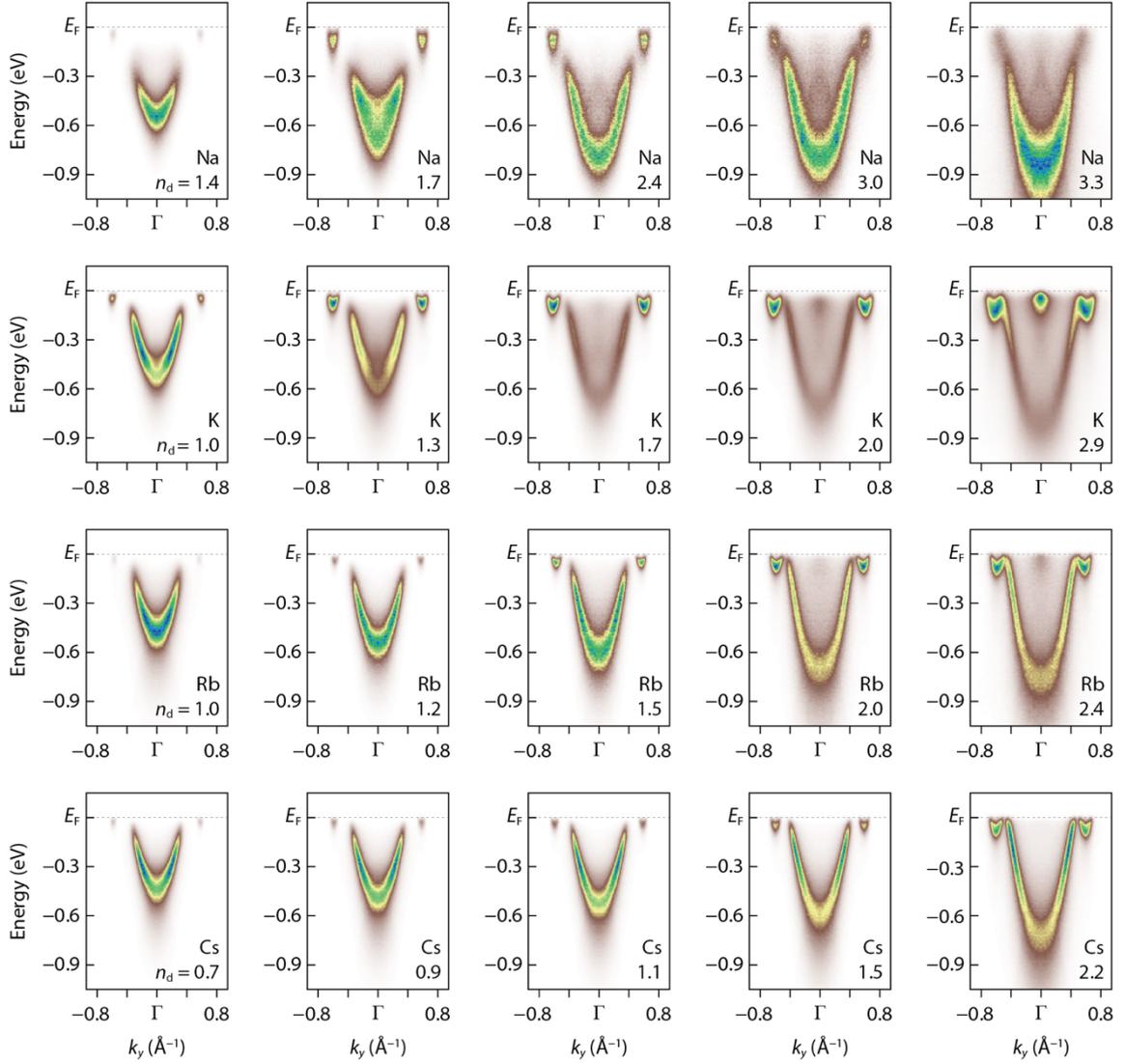

**Extended Data Figure 9 | Evolution of pseudogap with surface doping.** A doping series of APRES data is taken from black phosphorus whose surface is doped by Na, K, Rb, and Cs marked at the bottom right of each panel. Shown together in number is the dopant density $n_d$ in units of $10^{14}$ cm$^{-2}$. It can be clearly seen that the magnitude of pseudogap progressively reduces down to zero with increasing $n_d$, as summarized in the phase diagram (Fig. 5b). The hole-like states developing inside of the C1 band (pronounced for Na at $n_d = 3.3 \times 10^{14}$ cm$^{-2}$) are the valence band of black phosphorus[23]. The additional feature near $E_F$ and at the $\Gamma$ point (pronounced for K at $n_d = 2.9 \times 10^{14}$ cm$^{-2}$) is another conduction band of black phosphorus reproduced in first principles band calculations[24].

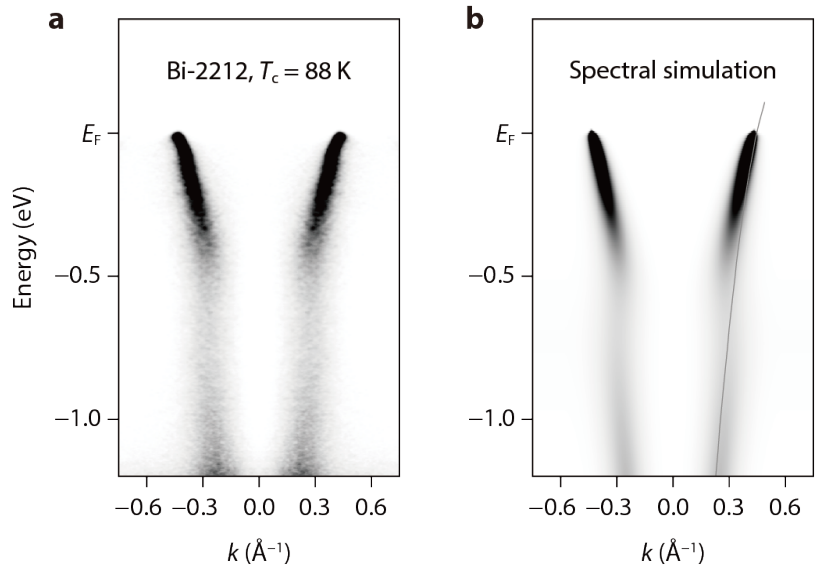

**Extended Data Figure 10 | Resonance effect for the waterfall dispersion in cuprates. a,** ARPES data taken from $Bi_{1.5}Pb_{0.5}Sr_2CaCu_2O_{8+\delta}$ ($T_c$ = 88 K) along the nodal direction with the photon energy of 98 eV. $n_d$ is estimated from $T_c$ (88 K)[35], pseudogap (32 meV)[36], and Luttinger count to be $1.2–1.4 \times 10^{14}$ cm$^{-2}$, corresponding to $k_r = 0.29 \pm 0.01$ Å$^{-1}$. **b,** Spectral simulations for resonance scattering at $k_r = 0.29$ Å$^{-1}$ (Methods). The grey curve represents the bare band. The waterfall dispersion in **a** is reproduced remarkably well by our spectral simulations in **b**.